# Modelling Localisation and Spatial Scaling of Constitutive Behaviour: a Kinematically Enriched Continuum Approach

**Giang Dinh Nguyen, Chi Thanh Nguyen, Vinh Phu Nguyen**
School of Civil, Environmental and Mining Engineering, University of Adelaide, Australia

**ABSTRACT**

It is well known that classical constitutive models fail to capture the post-peak material behaviour, due to localisation of deformation. In such cases the concept of Representative Volume Element (RVE) on which classical continuum models rest ceases to exist and hence the smearing out of local inhomogeneities over the whole RVE is no longer correct. This paper presents a new approach to capturing localised failure in quasi-brittle materials, focusing on the kinematic enrichment of the constitutive model to describe correctly the behaviour of a volume element with an embedded localisation band. The resulting models possess an intrinsic length scale which in this case is the width of the embedded localisation band. The behaviour therefore scales with both the width of the localisation band and the size of the volume on which the model is defined. As a consequence, size effects are automatically captured in addition to the model capability in capturing behaviour at the scale of the localisation zone.

## 1    INTRODUCTION

It is well known that failure in several engineering materials starts from a diffuse deformation process and gradually transforms to a localised deformation stage followed by material separation. Fracture Process Zone (FPZ) in quasi-brittle failure, shear bands in metals, and shear/compaction bands in soils are typical examples of localised failure. Constitutive models therefore have to correctly describe both diffuse and localised stages of failure. While it is straightforward to assume homogeneous deformation in diffuse stage of failure, such an assumption is no longer valid for a volume of material with an embedded localisation zone. In such cases, the homogeneity might be valid only at different scales of behaviour: homogenous deformation inside the localisation band, and in the region outside this band. In the literature, several different regularisation approaches have been proposed to tackle the modelling of localised failure. They range from simple (e.g. smeared crack/deformation [1]) to more complicated (nonlocal/gradient regularisation; [2, 3]), or focus on enriching the spatial discretisation schemes (e.g. eXtended Finite Element (XFEM [4]) and Enhanced Assumed Strain (EAS [5, 6]) to capture the kinematics of localised deformation.

Despite the success in describing localised failure, the above regularisations are either too simple to correctly capture the kinematics of localised deformation (smeared deformation), too complicated and computationally expensive to be practical applicable (nonlocal/gradient) or dependent on the discretisations (XFEM, EAS). In this study, the focus is on the constitutive modelling aspects of localised failure. This brings adequate enhancements to the kinematics of localised deformation while not making the proposed approach complicated or expensive and also facilitating its implementation in any spatial discretisation schemes.

The paper starts with a numerical illustration of failure in quasi-brittle materials. This shows the evolution of failure from diffuse to localised, and addresses associated issues in continuum modelling to capture this evolution. It is then followed by the introduction of a kinematically enriched continuum framework with different constitutive behaviour for the material inside and outside the localisation zone and their connections. Key features will be addressed, together with a numerical example to illustrate the potentials of the new approach.

## 2 LOCALISED FAILURE IN QUASI-BRITTLE MATERIALS

The development of micro-cracks together with the diffuse to localized failure of a RVE made of cement matrix composite is illustrated here. The RVE size is 25mm; plane stress condition (thickness 10mm) is assumed and the uniform mesh sizes are 0.25mm and 0.125mm for the coarser and finer meshes, respectively. The generation of sample is facilitated by the use of the Material Point Method (MPM [7]) and all three phases of the composite material including cement matrix (Young modulus $E$=45000MPa; uniaxial tensile strength $f_t'$=2MPa, fracture energy $G_F$=1.51Nmm/mm$^2$), inclusions ($E$=45000MPa, $f_t'$=3.0MPa, $G_F$=3.71Nmm/mm$^2$) and their weak interfacial transition zones ($E$=30000MPa, $f_t'$=1.5MPa, $G_F$=0.65Nmm/mm$^2$) can be generated using material points (Fig. 1). Further details on the modelling and constitutive models can be found in [8].

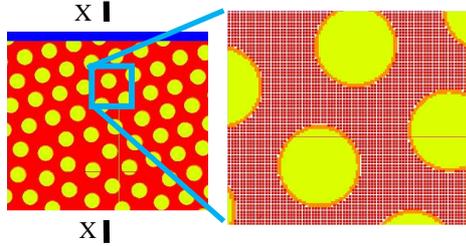
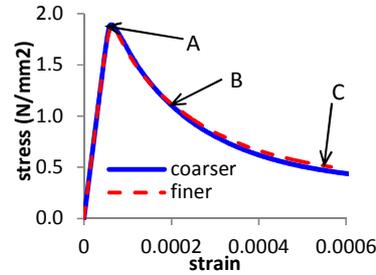

Figure 1: RVE made of cement matrix composite.

Figure 2: Mesh independent stress strain response.

The response of the RVE, in terms of macroscopic stress-strain relationship is plotted in Fig. 2, showing the softening branch associated with localised deformation inside the RVE. The micromechanical analysis in this case provides us with details on the distributions of stresses and strains during the failure process. For illustration purpose, only tensile loading in the vertical direction is considered.

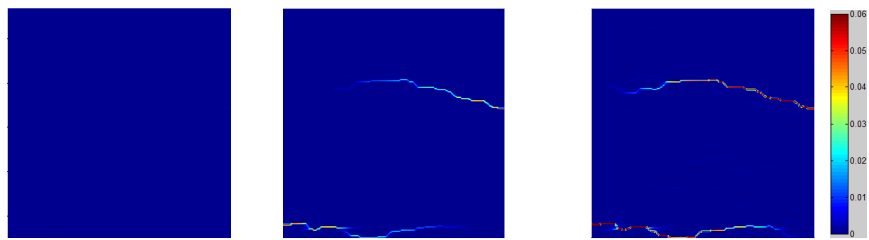

Figure 3: Contour of strain $\varepsilon_{yy}$ corresponding to points A, B, C in Fig. 2.

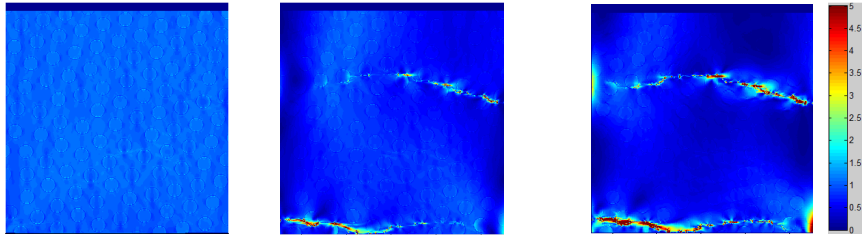
Figure 4: Contour of Von Mises stress corresponding to points A, B, C in Fig. 2.

In relation to the loading condition, the distribution of strain $\varepsilon_{yy}$ and Von Mises stress during failure are plotted in Figs. 3 & 4. It is clear that the distributions of both stress and strain are strongly localised towards the end of the failure process where macro crack and separation are expected. In this case continuum assumption breaks down as soon as the localisation of deformation is clearly visible, e.g. close to macro peak stress.

## 3   ENRICHING THE CONTINUUM KINEMATICS

From the above micromechanical analysis, it is clear that only a very small fraction of the RVE undergoes inelastic deformation during loading, while the rest just remain elastic all the time, or unloads elastically after peak. It is then desirable to capture as much as possible the micromechanical details and their evolution, while maintaining a balance between accuracy and practicality. In this sense a mechanism should be devised for mapping essential micromechanical details to a continuum model.

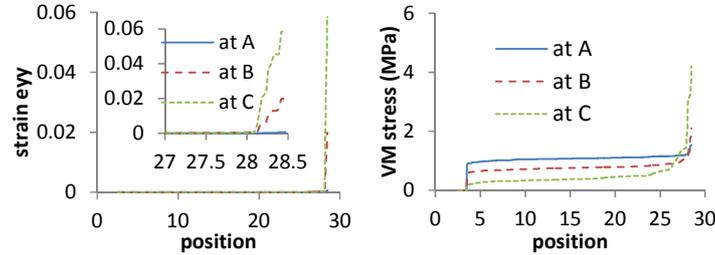
Figure 5: Strain and stress profiles on line X-X at different stages of failure in Fig. 2.

Given the spatial distribution of elastic and inelastic behaviour associated with different material points, two separate regions corresponding to elastic and inelastic behaviours in the RVE are needed in the continuum models. For illustration purpose, the Von Mises stresses, strains $\varepsilon_{yy}$ on the line X-X (Fig. 1) are sorted in ascending orders to give an indication of the evolution from diffuse to localised failure. This is presented in Fig. 5.

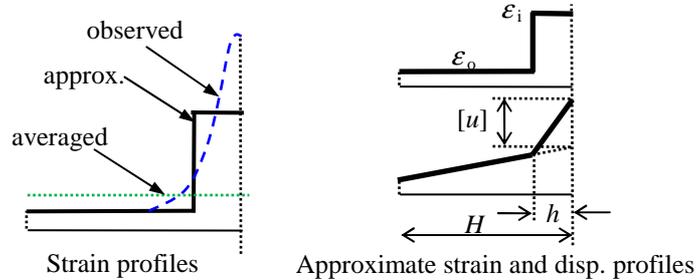
Figure 6: Strain profiles and approximation.

Due to the strong localisation of deformation, the use of a single-stress strain relationship is not appropriate, as an average strain profile is not a good representation of the observed behaviour (Fig. 6). A mechanism of localised failure is therefore needed, with the introduction of both stress $\boldsymbol{\sigma}_i$ and strain $\boldsymbol{\varepsilon}_i$ inside the localisation band to represent inelastic behaviour (Fig. 6). The elastic behaviour outside the localisation band is represented by a different stress-strain relationship $\boldsymbol{\sigma}_o$-$\boldsymbol{\varepsilon}_o$.

Homogeneity assumption is therefore valid only in each region (inside or outside the band; Fig. 6), and the macroscopic strain tensor is obtained from a simple homogenisation:

$$\dot{\boldsymbol{\varepsilon}} = f\dot{\boldsymbol{\varepsilon}}_k + (1-f)\dot{\boldsymbol{\varepsilon}}_o \qquad (1)$$

where $f$ is the volume fraction of the inelastic region [9]. Given the above strain, the macroscopic stress is obtained from the virtual work equation, so that:

$$\boldsymbol{\sigma}:\dot{\boldsymbol{\varepsilon}} = f\boldsymbol{\sigma}_i:\dot{\boldsymbol{\varepsilon}}_i + (1-f)\boldsymbol{\sigma}_o:\dot{\boldsymbol{\varepsilon}}_o \qquad (2)$$

For a narrow localisation band, e.g. $h \ll H$ (Figs. 5 & 6), the localisation strain $\boldsymbol{\varepsilon}_i$ can be assumed to take the following form (Kolymbas, 2009):

$$\dot{\boldsymbol{\varepsilon}}_i = \frac{1}{h}\left(\mathbf{n} \otimes [\dot{\mathbf{u}}]\right)^{\text{sym}} = \frac{1}{2h}\left(\mathbf{n} \otimes [\dot{\mathbf{u}}] + [\dot{\mathbf{u}}] \otimes \mathbf{n}\right) \qquad (3)$$

where $[\dot{\mathbf{u}}]$ is the relative velocity between opposite sides of the localisation band. In this case, (1-3) lead to $\boldsymbol{\sigma} \equiv \boldsymbol{\sigma}_o$ [9], and the constitutive equation describing the RVE behaviour becomes:

$$\dot{\boldsymbol{\sigma}} = \dot{\boldsymbol{\sigma}}_o = \mathbf{a}_o:\dot{\boldsymbol{\varepsilon}}_o = \frac{1}{1-f}\mathbf{a}_o:(\dot{\boldsymbol{\varepsilon}} - f\dot{\boldsymbol{\varepsilon}}_i) \qquad (4)$$

in which $\mathbf{a}_o$ is the elastic stiffness of the elastic region. Denoting $\mathbf{a}_i$ the tangent stiffness, the inelastic behaviour inside the localisation band is therefore:

$$\dot{\boldsymbol{\sigma}}_i = \mathbf{a}_i:\dot{\boldsymbol{\varepsilon}}_i = \frac{1}{h}\mathbf{a}_i:\left(\mathbf{n} \otimes [\dot{\mathbf{u}}]\right)^{\text{sym}} \qquad (5)$$

The internal equilibrium across the localisation band with normal vector $\mathbf{n}$ then completes the model definition:

$$(\dot{\boldsymbol{\sigma}} - \dot{\boldsymbol{\sigma}}_i) \cdot \mathbf{n} = 0 \qquad (6)$$

Equations (4-6) are generic structures of the constitutive equations describing the behaviour of a RVE crossed by a localisation band. The computational, energetic aspects, numerical implementation and examples demonstrating promising features of the kinematic enrichment of this type have been documented in some recent papers [9].

## 4    ONE-DIMENSIONAL NUMERICAL EXAMPLE

This numerical example shows the response of a bar of length $H$, with unit cross sectional area and an embedded localisation zone of size $h$. The material is assumed to be elastic-softening with linear softening law and specific fracture energy $g_F$ is taken as 20 times the elastic strain energy at peak, $g_F = 20 f_t^2/(2E)$. The critical bar length $L_c$ at which snap back starts to occur (e.g. $H > L_c$ for snap back) is therefore $L_c = 20h$. Fig. 7a shows the effects of varying the bar length $H$ on the overall response, while keeping the size $h$ of the localisation band fixed. The scaling of response with respect to size $H$ can be seen, while the energy dissipation produced by inelastic behaviour inside the localisation zone is insensitive to $H$, indicated by the same area under all three curves corresponding to three values of $H$. In all three cases, the constitutive response inside the localisation zone (Fig. 7b) is independent of the structural size $H$, thanks to the use of two separate constitutive laws for the localisation band and elastic region outside that band. In contrast, the single stress-strain law in the traditional smeared crack approach requires the variation of the constitutive behaviour with respect to the discretisation to meet the requirement on the dissipation (Fig. 7c).

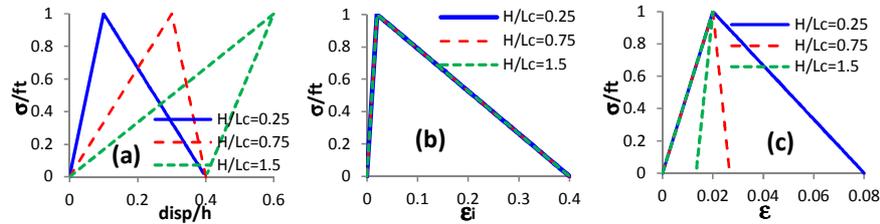

Figure 7: (a) normalised stress-normalised displacement; (b) normalised stress-strain response in the localisation zone of the proposed approach, (c) normalised stress-strain response of a smeared crack model.

The coupled stresses in the proposed constitutive modelling framework allow us to keep a meaningful constitutive response inside the localisation zone that is invariant with the discretisation (Fig. 7b). Essentially, all parameters of the model remain unchanged with respect to the resolution of the discretisation, a property that is missing in traditional smeared crack approach.

## 5  CONCLUSIONS

We used micromechanical analysis to identify key issues in constitutive modelling of materials that exhibit localised failure. An approach to enhance the constitutive kinematics was therefore proposed to capture correctly the material failure in both diffuse and localised stages. The resulting structure of constitutive models possesses both sizes and behaviours of the elastic and inelastic regions of a RVE crossed by a localisation band. The response of such models then scales with both the RVE size $H$, and the width $h$ of localisation band, while dissipation is dependent on $h$ and its associated inelastic behaviour only. Further work is on-going to apply this approach to the modelling of localised failure in geomaterials.

## 6  ACKNOWLEDGEMENTS

Support from the Australian Research Council (project DP140100945) is acknowledged.